\documentclass[unsortedaddress,preprint,aps,show pacs]{revtex4}
\usepackage{graphicx}
\usepackage{dcolumn}
\usepackage{amsmath}
\usepackage{amssymb}
\usepackage{amsfonts}
\usepackage{bm}
\usepackage{color}
\newcommand{\be}{\begin{equation}}
\newcommand{\ee}{\end{equation}}
\newcommand{\ba}{\begin{eqnarray}}
\newcommand{\ea}{\end{eqnarray}}
\begin{document}
\title{Hexatic phase and cluster crystals\\of two-dimensional GEM4 spheres}
\author{Santi Prestipino$^{1,2}$\footnote{Corresponding author. Email: {\tt sprestipino@unime.it}} and Franz Saija$^2$\footnote{Email: {\tt saija@ipcf.cnr.it}}}
\affiliation{$^1$Universit\`a degli Studi di Messina, Dipartimento di Fisica e di Scienze della Terra, Contrada Papardo, I-98166 Messina, Italy\\$^2$CNR-IPCF, Viale F. Stagno d'Alcontres 37, I-98158 Messina, Italy}
\date{\today}
\begin{abstract}
Two-dimensional crystals of classical particles are very peculiar in that melting may occur in two steps, in a continuous fashion, via an intermediate hexatic fluid phase exhibiting quasi-long-range orientational order. On the other hand, three-dimensional spheres repelling each other through a fast-decaying bounded potential of generalized-exponential shape (GEM4 potential) can undergo freezing into cluster crystals, allowing for more that one particle per lattice site. We hereby study the combined effect of low spatial dimensionality and extreme potential softness, by investigating the phase behavior of the two-dimensional (2D) GEM4 system. Using a combination of density-functional theory and numerical free-energy calculations, we show that the 2D GEM4 system displays one ordinary and several cluster triangular-crystal phases, and that only the ordinary crystal first melts into a hexatic phase. Upon heating, the difference between the various cluster crystals fades away, eventually leaving a single undifferentiated cluster phase with a pressure-modulated site occupancy. 
\end{abstract}
\maketitle

\section{Introduction}
The central paradigm of the statistical physics of simple liquids is that crystallization is essentially promoted by a harsh repulsion of quantum-mechanical origin between the atomic cores, as embedded e.g. in classical model interactions like the hard-sphere potential or the Lennard-Jones potential. The broad occurrence among atomic elements of highly-coordinated crystal structures like fcc and hcp gives evidence that considerations of packing efficiency are indeed relevant to solid formation also beyond the narrow bounds of rare-gas substances. However, optimality in space covering is arguably not the unique criterion behind crystal selection considering the widespread diffusion of bcc and lower-coordinated, even non-Bravais lattices among chemical elements, especially metals~\cite{Young}. If, on one hand, this would just indicate that short-range attractive forces also play a role, on the other hand open crystal structures are more often the effect of a {\em softening} of the interatomic repulsion at short distances, due generally to some pressure-induced reorganization in the electronic structure of the atoms~\cite{Jayaraman,Malescio,Prestipino1}. It is an amazing fact that the same kind of softness in the short-range repulsion between particles is realized in many colloidal fluids, made of complex macromolecules dispersed in a solvent, for reasons ultimately related to the multi-level architecture of their constituent molecules. An extreme form of softness is also possible, for example in self-avoiding polymers or in dissolved dendrimers~\cite{Louis,Lenz}, where the effective two-body repulsion keeps finite at the origin, allowing for full particle interpenetration.

In the last few years, the interest for simple fluids with a bounded interparticle potential has much revived after the observation made by Likos {\em et al.}~\cite{Likos1} that a new kind of crystals, called {\em cluster crystals}, can be made stable in these systems at high enough pressure. In a cluster crystal, more than one particle occupy the same crystal-lattice site (with the same average number $n_c$ of particles per blob or cluster) owing to the enthalpic prevalence, under sufficient pressure, of full particle overlap over more costly partial overlaps with many neighbors~\cite{Mladek1}. This cooperative effect is different from clustering in a gel-forming material~\cite{Pini,Sciortino,Bomont}, where clusters are rather stabilized by a combination of short-range attraction and long-range repulsion, resulting in liquid aggregates that disappear at high densities. Cluster crystals are rather exotic materials showing mass transport and an unusual reaction to compression~\cite{Mladek2} and shear~\cite{Nikoubashman}. However, no experimental realization of a cluster crystal has thus far been achieved, the correspondence with dendrimers being only observed at a modellistic level~\cite{Lenz}.

According to the criterion stated in Ref.~\cite{Likos2}, the condition for an isotropic bounded potential $u(r)$ to exhibit clustering at high pressure is a Fourier transform $\widetilde{u}(k)$ with both positive and negative parts. Among generalized-exponential, $\exp(-r^n)$ models, this requires $n>2$. The $n=4$ model, also known as the {\em generalized exponential model} of index 4 (GEM4), was extensively studied in three dimensions as a model for the effective pair repulsion between flexible dendrimers in a solution~\cite{Likos3,Mladek3,Mladek1}. The GEM4 system shows, among others, also an infinite number of cluster-crystal phases~\cite{Zhang1,Wilding1}. This is to be contrasted with the Gaussian case, $n=2$, where only two ordinary crystals (fcc and bcc) are stable~\cite{Stillinger,Prestipino2}.

In this paper we carry out the study of the GEM4 system in two dimensions. Ordinary (i.e., non-cluster) two-dimensional (2D) crystals are different from 3D crystals in that they cannot exhibit long-range positional order above zero temperature but only a weaker (quasi-long-range) form of translational order, characterized by an algebraic decay of positional correlations. This opens up the possibility that 2D melting occurs in two steps, via a so-called hexatic fluid phase with fairly extended (i.e., power-law decaying) bond-angle correlations, as originally predicted by the KTHNY theory~\cite{KTHNY}. In this respect, the question naturally arises as to whether also a 2D cluster crystal melts first into a hexatic-type fluid. The 2D GEM4 system offers an opportunity to study the interplay between clustering and bond-angle order in planar geometry.

After providing a proof of the existence of stable cluster crystals at zero temperature ($T=0$), we shall use Monte Carlo simulation to draw the complete phase diagram of the 2D GEM4, using the method introduced in Ref.\,\cite{Mladek3} for free-energy computations. We will see that a cluster-crystal phase emerges out of each stable $T=0$ crystal. In addition, a further ordinary crystal and a fluid phase are found at low pressure. A thin hexatic region is certainly present for low pressures, as an intermediate stage on the path from ordinary crystal to isotropic fluid, and a small equilibrium concentration of clusters will not alter this picture. On the other hand, full-blown clustering forces to rethink the definition of the orientational order parameter and, consequently, affects the very nature of the hexatic phase itself. One practicable solution would be to replace each cluster of particles with its center of mass (while isolated particles remain themselves). If we do this, we find that the first-order melting of the 2-cluster crystal occurs directly into the isotropic fluid, i.e., with no intermediate hexatic-like phase.

The outline of the paper is the following. After briefly recalling in Sect.\,II the definition of the model and the methods employed to study it in detail, we sketch the phase diagram in Sect.\,III, first by theoretical methods; then we show and discuss our simulation results in Sect.\,IV. Some concluding remarks are presented in Sect.\,V.

\section{Model and method}
\setcounter{equation}{0}
\renewcommand{\theequation}{2.\arabic{equation}}

The GEM4 potential is $u(r)=\epsilon\,\exp\{-(r/\sigma)^4\}$, where $r$ denotes the interparticle distance while $\epsilon>0$ and $\sigma$ are arbitrary energy and length units, respectively. In two dimensions, the GEM4 system can be described as a fluid of softly-repulsive disks of diameter $\sigma$, which can fully overlap with only a finite energy penalty ($\epsilon$, rather than infinity as in the case of hard disks). The Fourier transform of $u(r)$, which in 2D reads
\be
\widetilde{u}(q)=2\pi\int_0^\infty{\rm d}r\,ru(r)J_0(qr)\,\,\,\,\,\,{\rm with}\,\,\,\,\,\,J_0(x)=\frac{1}{2\pi}\int_0^{2\pi}{\rm d}\theta\cos\left(x\cos\theta\right)\,,
\label{2-1}
\ee
takes values of both signs ($J_0$ is a Bessel function of the first kind); hence, according to the criterion introduced in Ref.\,\cite{Likos2}, this system supports stable cluster phases at high pressure, as will later be confirmed by exact calculations at $T=0$ and by numerical simulations at $T>0$.

Before embarking on the simulation study of the 2D GEM4, it is convenient to first identify the relevant solid phases of the system. A way to do this is through the exact determination of the chemical potential $\mu$ as a function of the pressure $P$ at $T=0$. This task can be accomplished by means of exact total-energy calculations for a number of candidate perfect crystals. Precisely, for any given structure and value of $P$ we computed the minimum of $E+PV$ (where $E$ is energy and $V$ is volume) over a set of variables comprising the density $\rho=N/V$ and, possibly, also a number of internal parameters (see, e.g., Ref.\,\cite{Prestipino1}). On increasing the pressure, we find that the lowest chemical potential is sequentially provided by each $n$-cluster triangular crystal (dubbed ``$n$'' for clarity) of perfectly overlapping particles ($n=1,2,\ldots$); the transition from one state to another always occurs with a jump in the density, signaling first-order behavior (see Sect.\,III).

In order to investigate the phase diagram at high temperature, we used density-functional theory (DFT) in the mean-field (MF) approximation~\cite{Evans,Likos3}. DFT results for the 2D GEM4 will be presented in Sect.\,III. MF-DFT is very accurate in three dimensions~\cite{Mladek2,Zhang1}, at least for sufficiently high temperatures, hence we expect it to be reasonably effective also in two dimensions.

The system phase behavior at intermediate temperatures was investigated by Monte Carlo (MC) simulations in the $NPT$ (isothermal-isobaric) ensemble, with $N$ ranging from approximately 1000 up to 6048. The Metropolis algorithm, cell lists, and periodic boundary conditions were employed, as usual. Besides the standard local moves, in order to speed up the MC sampling an average 50\% of the displacements were directed towards randomly chosen positions in the simulation box. Typically, for each $(P,T)$ state point as many as $5\times 10^5$ cycles or sweeps (one sweep corresponding to $N$ trial MC moves) were generated at equilibrium, which proved to be enough to obtain high-precision statistical averages for the volume and the energy per particle. Much longer production runs of 2 million sweeps each were performed at closely-separated state points on a few isobars in order to study the melting of ``1'' in much more detail. 1 million sweeps at $P=0.2$ and $T=0.1$ were sufficient to accurately compute the chemical potential of the fluid phase by Widom's particle-insertion method~\cite{Widom}. The location of each phase transition was determined through thermodynamic integration of chemical-potential derivatives along isobaric and isothermal paths connecting the system of interest to a reference system whose free energy is already known (see, e.g., Ref.~\cite{Saija1}). While the reference state for the ``1'' phase was the triangular crystal at $P=0.5$ and $T=0.01$, deep inside the ``2'' region a low-temperature cluster crystal with occupancy 2 was taken as the starting point of our MC paths. In this state, the Helmholtz free energy was computed by a variant~\cite{Mladek2} of the Einstein-crystal method which is briefly described below. 

In a cluster crystal the average site occupancy $n_c$ is not a fixed parameter but it rather undergoes (usually slow) thermal relaxation like any other unconstrained variable. Its conjugate thermodynamic variable, $\mu_c$, a sort of chemical potential, spontaneously adjusts to zero in equilibrium~\cite{Swope,Mladek2}. This fact considerably complicates the numerical determination of the free energy with respect to ordinary crystals, since a further minimization of the free energy as a function of $n_c$ has to be carried out~\cite{Mladek2,Zhang1}.

In the original Frenkel-Ladd method~\cite{Frenkel,Polson}, the free energy of the system of interest is built up starting from the known free energy of a system of independent harmonic oscillators. To this aim, a linear morphing $U_\lambda$ ($0\le\lambda\le1$) between the two potential energies is introduced, and for each intermediate step $\lambda$ of the calculation the average energy difference $\langle\Delta U\rangle_\lambda$ between the two systems is computed in a $NVT$ simulation. In simulations of a crystal with multiply-occupied sites, a more appropriate choice of reference system is a gas of non-interacting particles diffusing in a landscape of disjoint potential barriers centered on the lattice sites~\cite{Mladek2}. As argued in Ref.\,\cite{Mladek3}, in order to improve the sampling efficiency for $\lambda=1$ (corresponding to $U_\lambda=U$, i.e., the actual potential energy) on average one MC move out of $N$ was attempted to shift the system center of mass to a totally random position in the box. For other values of $\lambda$, the trial moves were of the same type as considered for $NPT$ runs. Particular care was paid in the extraction of $F$ from $\langle\Delta U\rangle_\lambda$, since the latter quantity exhibits a strong dependence on $\lambda$ near 0 and 1.

Clearly, by the above outlined method only the Helmholtz free energy $F(n_c)$ for a preset value of $n_c$ {\em and} a specific state point is computed. In principle, at each $(P,T)$ point one should carry out the free-energy calculation for several $n_c$'s and eventually pick the one with the minimum associated $\mu$, but this is obviously a rather daunting task. Rather efficient methods to do the calculation of the equilibrium value of $n_c$ have actually been developed\,\cite{Zhang2,Wilding2}, but we shall anyway follow a simpler, even though more tedious procedure. In principle, it would be sufficient to compute $\mu$ at one reference state for many $n_c$ values and then generate a separate (either isobaric or isothermal) chain of Monte Carlo runs for each of them. The actual system chemical-potential curve along a given MC path will be the lower envelope of the various $\mu(n_c)$ curves along that path. This is a legitimate procedure which, however, is doomed to fail if $n_c$ is not conserved along the respective path. Even in case of an exactly conserved non-integer $n_c$, one should nonetheless check at each state point -- by, e.g., visual inspection on a random basis -- that the different integer occupancies are randomly distributed among the sites. 

\section{Zero-temperature behavior and DFT analysis}
\setcounter{equation}{0}
\renewcommand{\theequation}{3.\arabic{equation}}

We first looked at the $T=0$ phases of the system. Assuming periodic structures only, we initially examined all the Bravais lattices (there are five of them in 2D) and the honeycomb lattice. Strictly periodic configurations permit the exact computation of the system chemical potential, reducing it to seeking the minimum of a few-parameter expression. Except for the oblique lattice, which has two internal parameters to optimize, we put an equal number of particles on each lattice site, up to five. As anticipated in Sect.\,II, it turns out that only $n$-cluster triangular crystals ($n=1,2,\ldots$) provide stable phases of the 2D GEM4 at $T=0$. For low enough pressures, the ordinary ($n_c=1$ or ``1'') crystal provides the minimum chemical potential. The ``1'' crystal is eventually superseded at $P=0.7236$ (reduced units) by the 2-cluster crystal ($n_c=2$ or ``2'' crystal), which represents the most stable phase up to $P=2.1306$; and so on (see Table 1).

%
%
\begin{table}
\caption{Zero-temperature phases of the 2D GEM4 system (only data the first four phases are listed). For each pressure range in column 1, the thermodynamically stable phase is indicated in column 3, together with the respective values of the number density $\rho$ (column 2).}
\begin{tabular*}{\columnwidth}[c]{@{\extracolsep{\fill}}|c|c|c|}
\hline
$P$ range ($\epsilon/\sigma^2$) & $\rho$ range ($\sigma^{-2}$) & stable phase \\
\hline\hline
0-0.7235 & 0-0.69399 & ``1'' ($n_c=1$) \\
\hline
0.7236-2.1306 & 1.10292-1.30386 & ``2'' ($n_c=2$) \\
\hline
2.1307-4.2415 & 1.71512-1.91210 & ``3'' ($n_c=3$) \\
\hline
4.2416-7.0560 & 2.32394-2.51964 & ``4'' ($n_c=4$) \\
\hline
\end{tabular*}
\end{table}

We might reasonably expect that the sequence of stable phases remains unchanged for sufficiently small non-zero temperatures. However, we soon realize that other states may arise at $T>0$, different from triangular cluster crystals with integer $n_c$, especially near the transition pressures listed in Table 1.

As a matter of example, in a range of pressures around $P=0.7236$ we have computed the $T=0$ chemical potential for a number of periodic structures interpolating between ``1'' and ``2'', i.e., containing single particles and pairs in various combinations (for simplicity, only perfect triangular crystals have been considered, see Fig.\,1). For instance, in the triangular lattice denoted ``12a'', isolated particles (1's) occur regularly alternated with pairs of fully overlapping particles (2's). Therefore, this lattice has $n_c=1.5$. The lattice denoted ``12b'' has the same value of $n_c$ but the pairs are now lined up along one principal lattice direction. The chemical-potential data for the structures in Fig.\,1 are reported in the right panels of Fig.\,2 (note that, in all figure labels, we restore absolute units for $P,T,\mu$, etc.). We see that all these crystals acquire nearly the same $\mu$ of ``1'' and ``2'' for $P\simeq 0.7236$. What is amazing is that crystal states with widely different site occupancies are nearly degenerate at these pressures, suggesting a non-trivial transition scenario for $T>0$ characterized by a slow relaxation dynamics, very much like that found in 1D~\cite{Prestipino3}. Moreover, periodic configurations characterized by an equal $n_c$ but different spatial distribution of 1's and 2's have almost the same $\mu$. That said, it also appears that, on approaching $P=0.7236$ from below, the statistical weight of a configuration of 1's and 2's would be larger the closer its $n_c$ value is to 2, implying that, as $T$ grows, the coexistence value of $n_c$ increasingly drifts away from 2, exactly as found in 3D~\cite{Zhang1}, due ultimately to the entropic advantage of any $n_c<2$ over $n_c=2$. Therefore we expect that the equilibrium $n_c$ value for ``$n$'' ($n=1,2,\ldots$) acquires a pressure dependence for $T>0$, reducing to $n$ only far away from phase boundaries and for not too high temperatures.

On the opposite side of the temperature axis, accurate results on the phase boundaries of the system can be obtained from MF-DFT~\cite{Likos3}. In presenting this theory, we mainly follow Ref.\,\cite{Prestipino4}, though we note that our results for the 3D system~\cite{Prestipino3} are substantially equivalent to those reported in Ref.\,\cite{Likos3}, where a slightly different formalism had been developed -- built around the $NVT$ ensemble, rather than the $\mu VT$ ensemble employed here.

In 2D, the grand-potential functional $\Omega_\mu[n]$ of the crystal one-point density $n({\bf x})$ is written as
\be
\Omega_\mu[n]=F^{\rm id}[n]+F^{\rm exc}[n]-\mu\int{\rm d}^2x\,n({\bf x})\,,
\label{3-1}
\ee
with separate, ideal (known) and excess (generally unknown) contributions to the Helmholtz free-energy functional $F[n]$. In the Ramakrishnan-Yussouff theory~\cite{Ramakrishnan}, the two-point direct correlation function (DCF) of the crystal is approximated with that, $c(r;\rho)$, of the homogeneous fluid of density $\rho$, thus leading to an excess free energy of
\be
\beta F^{\rm exc}[n]=\beta F^{\rm exc}(\rho)-c_1(\rho)\int{\rm d}^2x(n({\bf x})-\rho)-\frac{1}{2}\int{\rm d}^2x\,{\rm d}^2x'\,c(|{\bf x}-{\bf x}'|;\rho)(n({\bf x})-\rho)(n({\bf x}')-\rho)\,,
\label{3-2}
\ee
where $c_1(\rho)=\ln(\rho\Lambda^2)-\beta\mu$ is the one-point DCF of the fluid, $\Lambda$ is the thermal wavelength, and $F^{\rm exc}(\rho)$ is the excess free energy of the fluid. A further simplifying, MF-like assumption is $c(r)\approx-\beta u(r)$, which eventually gives $F^{\rm exc}$ the compact form
\be
F^{\rm exc}[n]=\frac{1}{2}\int{\rm d}^2x\,{\rm d}^2x'\,u(|{\bf x}-{\bf x}'|)n({\bf x})n({\bf x}')\,.
\label{3-3}
\ee

Now let $M$ and $N_s$ denote the number of lattice sites and the average number of solid particles, respectively. In terms of the solid density $\rho_s=N_s/V$ and the unit-cell volume $v_0=V/M$, the average site occupancy reads $n_c=\rho_sv_0$. A popular {\em ansatz} for $n({\bf x})$ is
\be
n({\bf x})=n_c\left(\frac{\alpha}{\pi}\right)\sum_{\bf R}e^{-\alpha({\bf x}-{\bf R})^2}=\rho_s\sum_{\bf G}e^{-G^2/(4\alpha)}e^{i{\bf G}\cdot{\bf x}}\,,
\label{3-4}
\ee
where the last equality transforms a direct-lattice sum into a reciprocal-lattice sum. Plugging Eq.\,(\ref{3-4}) into (\ref{3-1}), one obtains the grand-potential difference $\Delta\Omega_\mu[n]=\Omega_\mu[n]-\Omega(\rho)$ between the crystal and the fluid as a function of three parameters ($v_0,\alpha$, and $n_c$). At equilibrium, this function has to be made minimum for each $\rho$. For the given $\mu,V,T$, the freezing transition occurs where the minimum $\Delta\Omega_\mu$ ($\equiv\Delta\Omega^*$) happens to be zero. An explicit expression of $\Delta\Omega_\mu[n]$ is
\be
\frac{\beta\Delta\Omega_\mu}{V}=\frac{1}{V}\int{\rm d}^2x\,n({\bf x})\ln\frac{n({\bf x})}{\rho}-(\rho_s-\rho)+\frac{\beta}{2}(\rho_s-\rho)^2\widetilde{u}(0)+\frac{\beta}{2}\rho_s^2\sum_{{\bf G}\ne 0}e^{-G^2/(2\alpha)}\widetilde{u}({\bf G})\,.
\label{3-5}
\ee
The high-precision evaluation of the integral in Eq.\,(\ref{3-5}) can be made by the method explained in \cite{Wallace} (see also Ref.\,\cite{Prestipino5}).

From the general formulae
\be
c_1'(\rho)=\int{\rm d}^2x\,c(|{\bf x}-{\bf x}'|;\rho)=\widetilde{c}(0;\rho)\,\,\,\,\,\,{\rm and}\,\,\,\,\,\,\beta f^{\rm exc}(\rho)=-\frac{1}{\rho}\int_0^\rho{\rm d}\rho'(\rho-\rho')\widetilde{c}(0;\rho')\,,
\label{3-6}
\ee
it follows in the mean-field approximation that
\be
c_1(\rho)=-\beta\widetilde{u}(0)\rho\,\,\,\,\,\,{\rm and}\,\,\,\,\,\,\beta f^{\rm exc}(\rho)=\frac{\beta\widetilde{u}(0)}{2}\rho\,.
\label{3-7}
\ee
Hence, the solid grand potential can be written as
\be
\frac{\beta\Omega[n]}{V}=\frac{\beta\Delta\Omega[n]}{V}-\rho-\frac{\beta\widetilde{u}(0)}{2}\rho^2\,.
\label{3-8}
\ee
The fluid and solid equilibrium pressures are finally given in terms of the fluid density $\rho$ by
\be
\beta P_{\rm fluid}=-\frac{\beta\Omega(\rho)}{V}=\rho+\frac{\beta\widetilde{u}(0)}{2}\rho^2\,\,\,\,\,\,{\rm and}\,\,\,\,\,\,\beta P_{\rm solid}=\beta P_{\rm fluid}-\frac{\beta\Delta\Omega^*(\rho)}{V}\,.
\label{3-9}
\ee

A few results for the 2D GEM4 system are reported in Fig.\,3 and the resulting phase diagram is shown in Fig.\,4. On the basis of the MF-DFT theory, for all temperatures down to $T=0.1$ a first-order phase transition from fluid to cluster crystal is predicted. Within the cluster-crystal region, the density $\rho_s$, the $\alpha$ value, and the average site occupancy $n_c$ all increase almost linearly with pressure.

\section{Simulation results}
\setcounter{equation}{0}
\renewcommand{\theequation}{4.\arabic{equation}}

The exact calculations of Sect.\,III only apply for $T=0$, and were useful to identify the solid phases that are relevant for the 2D GEM4 system at $T>0$. The rest of our analysis was carried out by MC simulation.

In order to gain insight into the melting behavior at low pressure (say, for $P<0.70$), we performed a series of concatenated $NPT$ runs along a number of isobaric paths, first in large steps of $\Delta T=0.005$, starting from the triangular crystal at $T=0.01$. On the high-$T$ side, we descended in temperature from the fluid at $T=0.1$. The last configuration produced in a given run is taken to be the first of the next run at a slightly different temperature. Isothermal paths were generated at $T=0.01,0.02,\ldots,0.06$ in order to trace the coexistence line of the low-pressure phases (ordinary crystal and fluid) with the 2-cluster phase, taking $n_c=2$ cluster crystals at $P=1$ as starting points of the cluster-solid-type trajectories. In these simulations the pressure was changed in steps of $\Delta P=0.02$. We typically employed $5\times 10^5$ MC sweeps to compute the equilibrium averages at each state point, after discarding an equal number of sweeps in order to equilibrate the system from the previous point on the path. Closer to the transition point, we reduced the temperature and the pressure steps while leaving the rest of the simulation schedule unchanged.

In Fig.\,5 the system density $\rho$ is plotted as a function of the temperature, for a few $P$ values ranging from $0.1$ to $0.5$ (at these pressures, no fully overlapping particles were ever observed). As neatly shown by these graphs, upon heating the system melts continuously (i.e., no hysteresis is observed), showing exactly the same behavior as found in the 2D Gaussian case~\cite{Prestipino6} (the energy per particle, not shown, is itself a smoothly increasing function of $T$ across the transition). At $P=0.5$, we checked the order of the melting transition independently through thermodynamic integration combined with exact free-energy calculations, confirming second-order behavior (see Fig.\,5 inset). In the small $T$ interval where the rate of density change is higher the system phase is in all probability hexatic, as corroborated by the behavior of the orientational correlation function (OCF) $h_6(r)$ along the solid-type trajectories (see, e.g., Fig.\,6 for $P=0.5$ -- the precise OCF definition can be found in Ref.\,~\cite{Prestipino6}).

At the three pressures of Fig.\,5 the fluid is less dense than the crystal, implying that the melting line is increasing, at least up to $P=0.5$. However, the density difference between the ordinary crystal and the isotropic fluid reduces as the pressure grows, until its sign eventually reverses at $P\lesssim 0.65$, signaling a reentrant-melting behavior for higher pressures, again similarly to the Gaussian case (the maximum melting temperature is $T_M\simeq 0.044$). At these high pressures the system behavior upon heating is however more complicate, due the appearance of 2-particle clusters in successive waves, making the crystal density of the finite system a piecewise continuous function of $T$ (see more on this point below). For pressures in the reentrant-melting region the fluid density attains a shallow maximum upon isobaric cooling (for example at $T=0.071$ and $T=0.090$ for $P=0.70$ and $P=0.75$, respectively), indicating waterlike anomalous volumetric behavior.

We now report on the system behavior along a few $T>0$ isotherms, with a look at the transition between the ordinary crystal (``1'') and the 2-cluster crystal (``2''). We first used thermodynamic integration to draw the chemical potential $\mu$ of the crystals with $n_c=1$ and $n_c=2$ as a function of $P$, seeking for the point where $\mu_1(P)$ and $\mu_2(P)$ intersect each other. However, the locus of these intersections is only a crude estimate of the true transition line, the more so the higher $T$ is, since the equilibrium $n_c$ value of ``1'' and ``2'' is integer (1 and 2, respectively) only far away from transition points~\cite{Zhang1}; therefore, in order to locate the exact low-pressure boundary of the ``2'' phase it would be necessary to compute $n_c$ state by state. For simplicity, we will satisfy ourselves with a rough analysis, using the method described at the end of Sect.\,II. Therefore, below $P=1$ we attempted to compute the chemical potential of the $n_c=1.8$ and $n_c=1.9$ crystals (in the stable ``1'' phase, $n_c$ is practically 1 up to $T=0.03$; above this temperature, it spontaneously adjusts to equilibrium during the simulation). For $1<n_c<2$, the starting configuration of the simulation was chosen by randomly mixing, in the correct proportions, single particles with fully overlapping pairs. At the lowest probed temperature of $T=0.01$, it turns out that the $n_c=1.8$ crystal is never stable for $P<1$, since either the system eventually phase-separated in the course of the simulation into $n_c=1$ and $n_c=2$ crystals or the average site occupancy drifted away from 1.8 (see Fig.\,7). None of these problems occurred for $n_c=1.9$, at least at and below $P=0.8$, and we were then able to compare its chemical potential with that of $n_c=2$ near the ``1''-``2'' transition boundary. Looking at Fig.\,8 (top left panel), we conclude that at $T=0.01$ the value of $n_c$ stays close to 2 down to the transition point.

Along the isobar $P=1$ we were able to stabilize the $n_c=1.8$ crystal only in a small $T$ interval around $T=0.04$, whereas the identity of the $n_c=1.9$ crystal was preserved also above this temperature. On heating from $T=0$, the equilibrium value of $n_c$ gradually reduces from the initial 2 to roughly 1.9 at $T=0.06$, see Fig.\,8 top right panel.

We could compare the chemical potential of the three cluster crystals with $n_c=1.8,1.9$, and 2 with that of ``1'' only for $T=0.04$. As illustrated in the bottom left panel of Fig.\,8, along this isotherm the average occupancy of ``2'' progressively lowers on approaching ``1'', reaching approximately 1.8 at the transition point. Note, however, that the ``1''-``2'' coexistence at $T=0.04$ is probably only a metastable one, since the ``2'' phase apparently melts on decompression before transforming into ``1'' (see the next Figs.\,9 and 10). Finally, we illustrate the $T=0.06$ case in the bottom right panel of Fig.\,8. At this temperature, $n_c\approx 1.9$ at the (melting) transition.

Putting all things together, the (low-pressure, low-temperature) phase diagram of the 2D GEM4 system is reported in Fig.\,9, whereas the system phase diagram on the $\rho$-$T$ plane is depicted in Fig.\,10. Comparing the numerical $P$-$T$ phase diagram with the MF-DFT phase diagram, we conjecture that the melting line of the 2-cluster phase merges at high $T$ with the transition line predicted by DFT. Indeed, the DFT transition point at $T=0.1$ is roughly aligned with the boundary in Fig.\,9 separating the fluid phase from the 2-cluster phase. Along the low-pressure boundary of the 2-phase, the value of $n_c$ would then attain a minimum of roughly 1.8 at the triple point, $T_t\lesssim 0.040$, growing more rapidly above this temperature. As far as the other cluster phases are concerned, we did not carry out any systematic study at $T>0$ but we do expect to find a behavior similar to what found in 3D, where the coexistence line between each pair of successive cluster phases terminates with an Ising critical point~\cite{Zhang3}. On increasing the temperature, the transitions from ``$n$'' to ``$n+1$'' would become rounded one after another and it will then be possible to go continuously from one cluster phase to another simply by circumnavigating around the critical points, which again is consistent with the picture emerged from DFT of an undifferentiated cluster phase of high temperature whose average occupancy smoothly increases with $P$.

We finally discuss the nature of the melting transition of the ``1'' phase for pressures above $P=0.65$. We carried out MC production runs of $5\times 10^5$ to $2\times 10^6$ sweeps each, for two different system sizes ($N=1152$ and $N=2688$). In Fig.\,11 the system density $\rho$ and the specific energy $e=E/N$ are plotted as functions of $T$ along the $P=0.7$ isobar, i.e., very close to the low-pressure boundary of the ``2'' phase. First looking at the crystal, both $\rho$ and $e$ show a step-like behavior on heating, each step or jump being associated with the appearance of a new bunch of overlapping particles in the system. Each new step probably occurs after the triangular array has been completely restored from the last avalanche of cancelled sites. Once formed, these bound pairs would behave as quenched defects in the sea of 1's, as their lifetime is at least comparable to our observation time. This view is substantiated by the thermal evolution of the radial distribution function (RDF), whose first peak (i.e., the one centered at zero) grows itself by successive finite increments, see Fig.\,12. Within the ``1'' phase, the average site occupancy $n_c$, i.e., the average ratio of the particle number $N$ to the number of ``sites'' $N_s$, can be estimated as follows: rather than computing $N_s$ in each system configuration, we look at the average number $N(r<r_0)$ of particles whose distance from a reference particle at the origin is less than the width ($r_0$) of the first RDF peak. Since for each given configuration the number of overlapping particle pairs is $N-N_s$, it is $N-2(N-N_s)=2N_s-N$ the number of unpaired or isolated particles, and we get:
\ba
2\pi\rho\int_0^{r_0}{\rm d}r\,rg(r)\equiv N(r<r_0)&=&\frac{2\langle N_s\rangle-N}{N}\times 0+\left(1-\frac{2\langle N_s\rangle-N}{N}\right)\times 1
\nonumber \\
&=&2\left(1-\frac{\langle N_s\rangle}{N}\right)\approx 2-\frac{2}{n_c}\,,
\label{4-1}
\ea
assuming that the distribution of $N_s$ is sharply peaked. Using the above formula, the values of $n_c$ along the $P=0.7$ path have been reported in the inset of Fig.\,12, showing a behavior similar to that of the solid density. At each temperature, the average number of particles involved in clusters is approximately $2N(1-1/n_c)$.

We see from Fig.\,11 that $\rho$ and $e$ exhibit a step-like behavior also along the fluid-type trajectory. However, at variance with the crystal case, the fluid density and energy only show a single step down, followed for lower $T$ by an evolution resembling that of a crystal with long-lasting line defects. Therefore, the point where the density jump occurs gives a rather precise estimate of the freezing temperature. The complete healing of the transformed system from the residual crystal defects requires a time much longer than our typical simulation time, hence there would be no chance to observe it during our runs.

Rather than an undesired occurrence of our simulations, the onset of particle pairing on heating above $P=0.65$ is admittedly an intrinsic trait of the ``1'' equilibrium behavior, as suggested by the fact that the $\rho$ and $e$ jumps are more numerous and smaller for the larger of our samples. As the thermodynamic limit is approached, the sequence of jumps would eventually be converted to a smooth increase, thus bridging the gap between the cold-crystal and fluid branches of $\rho$ and $e$. Hence the melting transition of ``1'' will be continuous also along the reentrant-melting branch, while it is hard to say whether the hexatic phase is stable for these pressures (consider that, for a particle belonging to a cluster, the usual definition of the orientational order parameter~\cite{Prestipino6} does not apply). However, we argue that the 2-particle clusters in the system act as pinned defects for the triangular array, and we know that the stability of the hexatic phase gets {\em enhanced} (rather than suppressed) by the presence of quenched disorder~\cite{Deutschlaender}. Therefore, we surmise that the ``1'' melting is KTHNY-like also along the reentrant branch.

Different would be the case of the melting of ``2'' at, say, $T=0.06$. Here the phase transition is only weakly first-order, which leaves open the possibility of an unusual melting scenario for a cluster crystal, at least for not too high temperatures. However, since the particles taking part of clusters are now the vast majority, we are obliged to think over the definitions of bond angles and OCF. One possibility would be to look at the bonds between the centers of mass of neighboring clusters and the correlations between their orientations in space. We may then imagine a thermodynamic regime where the large-distance decay of bond-angle correlations is algebraic rather than just exponential, and this will entail a new kind of hexatic-type order. Checking the relevance of this scenario for the 2-cluster crystal of the 2D GEM4 is not easy, since large system sizes and long simulation times would be needed. We have anyway considered the melting behavior of a small-sized cluster crystal with $n_c=1.9$, starting from the equilibrated crystal at $P=0.9$ and $T=0.04$ and then heating up the system initially in steps of $\Delta T=0.001$ at constant pressure. Two distinct triangular-lattice sizes were considered, $24\times 28$ ($N=1276$) and $48\times 56$ ($N=5107$). For the smaller sample, $5\times 10^5$ sweeps were generated at each state point to equilibrate the system from the previous run at a slightly smaller temperature, and a further $5\times 10^5$ sweeps were considered for the calculation of the thermal averages (the numbers of MC cycles produced for the larger sample were four times larger). At regular intervals, we checked that single-particle defects are randomly distributed within the crystal. For both samples, we found that the cluster crystal melts abruptly into the isotropic fluid (Fig.\,13), with no evidence of an intermediate regime of power-law decay of angular correlations (see the OCF evolution in Fig.\,14). Whence we conclude that the melting of a 2D cluster crystal is almost certainly conventional 3D-like, at least in the present GEM4 model.

\section{Conclusions}

In the present study, the thermodynamic behavior of the 2D GEM4 system was investigated by DFT and MC simulation. Specialized numerical free-energy methods were employed in order to map out the equilibrium phase diagram of the system, thus unvealing an endless sequence of cluster-crystal phases in addition to a low-density ordinary crystal and a fluid phase.

We concentrated on the melting behavior of the 2D GEM4 system at low pressure, where simulation is less affected by the complications deriving from a slow relaxation to equilibrium. While the melting transition occurs continuously for the ordinary crystal, through an intermediate stage having all the characteristics of the hexatic phase, it is seemingly discontinuous for the cluster phase of lowest pressure (2-cluster crystal). Despite this, an unconventional melting behavior, involving a non-trivial interplay between clustering and bond-angle order, is conceivable for a 2D cluster crystal, at least provided that the notion of orientational order parameter is reformulated with reference to bonds between neighboring clusters rather than particles. If we adopt this modified measure of orientational correlations, we discover that the sudden isobaric melting of the 2-cluster crystal brings the system directly into the isotropic fluid (i.e., there is no trace of an intermediate temperature regime characterized by the algebraic decay of bond-angle correlations). As the temperature grows, the distinction between the various cluster phases becomes more vague till completely vanishing. At very high temperature a unique cluster phase will be left, whose average occupancy increases smoothly and almost linearly with pressure. The freezing of the fluid into this phase is, according to MF-DFT, strongly first-order.

\newpage
\section*{Figure captions}
%
%
{\bf FIGURE 1}. (Color online). Schematic of the 2D periodic structures, interpolating between ``1'' and ``2'', whose chemical potential was computed at $T=0$ (see it plotted as a function of the pressure in the bottom right panel of Fig.\,2). Single particles (1's) are marked by a cyan dot, whereas pairs of perfectly overlapping particles (2's) are depicted as a red dot.

%
%
{\bf FIGURE 2}. (Color online). Zero-temperature chemical potential of various 2D crystals relative to that of the ordinary triangular crystal (``1''), $\Delta\mu=\mu-\mu_1$, plotted as a function of the pressure $P$ (see main text and Fig.\,1 for notation). Solid lines: ``2'', blue; ``3'', cyan; ``4'', magenta; ``5'', red; rectangular, yellow; centered rectangular, green; oblique, black. Left-panel dotted lines: honeycomb, black; 2-cluster honeycomb, blue; 3-cluster honeycomb, cyan. Left-panel dashed lines: square, black; 2-cluster square, blue; 3-cluster square, cyan. Right-panels dotted lines: ``12a'', black; ``112a'', red; ``122a'', cyan; ``1222a'', blue. Right-panels dashed lines: ``12b'', black; ``112b'', red; ``122b'', cyan; ``1222b'', blue. The vertical dotted lines mark the transition pressures (see Table 1).

%
%
{\bf FIGURE 3}. (Color online). MF-DFT results for the 2D GEM4 system. The data, which refer to a number of temperatures (from left to right, $T=0.1,0.3,0.5,1$), are plotted as a function of the {\it reference-fluid density} $\rho$. For each $T$, the vertical dotted line marks the phase transition from fluid to cluster crystal. Top left panel: grand-potential difference between solid and fluid. Top right panel: solid density. Bottom left panel: $\alpha$ parameter. Bottom right panel: average site occupancy.

%
%
{\bf FIGURE 4}. MF-DFT for the 2D GEM4 system. Top: $P$-$T$ phase diagram. Bottom: $\rho$-$T$ phase diagram.

%
%
{\bf FIGURE 5}. (Color online). Number density $\rho$ of the 2D GEM4 system along the three isobars $P=0.1$ (left), $P=0.3$ (center), and $P=0.5$ (right). Data for various sizes ($N=2688$, open symbols; $N=6048$, full symbols) and simulation trajectories (solid-like, red triangles; fluid-like, blue dots) are shown. For the $N=6048$ system, $10^6$ sweeps were produced in each production run. No sign of hysteresis is observed at the three pressures considered, implying a continuous melting transition. Inset: chemical potential of the ordinary-crystal (red) and fluid (blue) phases for $N=1152$.

%
%
{\bf FIGURE 6}. (Color online). Orientational correlation function $h_6(r)$ of the 2D GEM4 system in the transition region for $P=0.5$ and two sizes, $N=2688$ (left) and $N=6048$ (right). Top panels: log-log scale. Bottom panels: log-linear scale. As $T$ increases, the large-distance behavior of $h_6(r)$ changes from constant (triangular solid) to power-law decay (hexatic fluid), eventually to exponential decay (isotropic fluid). According to the KTHNY theory, in the hexatic region the decay exponent should be less than $1/4$ (the slope of the black dashed curve).

%
%
{\bf FIGURE 7}. (Color online). A few snapshots taken from our $T=0.01$ simulations of a GEM4 system originally prepared in a $n_c=1.8$ state (top panels) or in a $n_c=1.9$ state (bottom panels), with an initially random distribution of clusters. Left: $P=1$ (isolated particles, cyan dots; particles belonging to clusters, red dots). Right: $P=0.78$. Note the differences existing between the cases represented in the top and bottom panels, respectively. Top: while $n_c$, computed through Eq.\,(\ref{4-1}), stayed fixed at 1.8 for all the duration of the run made for $P=1$, it changed irreversibly along the simulation path ending at $P=0.78$, eventually settling at a value slightly larger than 1.91. Bottom: again, the system phase-separates into a mixture of the $n_c=1$ and $n_c=2$ crystals at $P=1$, whereas below $P\simeq 0.80$ the mixing of 1's and 2's is totally random. Note the occurrence of a small number of vacancies in the snapshots taken for $n_c=1.9$ and $T=0.78$. These vacancies apparently occur at random positions in the underlying lattice; they may contribute to speed up the jump dynamics of particles in a cluster crystal, thus providing one mechanism (the main one?) for mass transport within such solids.

%
%
{\bf FIGURE 8}. (Color online). Chemical potential $\mu$ of the 2D GEM4 system for various phases along a number of paths on the $(P,T)$ plane (``1'', black solid line; $n_c=2$ crystal, red solid line; $n_c=1.9$ crystal, red dashed line; $n_c=1.8$ crystal, red dotted line; fluid, blue solid line). Top left: $T=0.01$. Top right: $P=1$. Bottom left: $T=0.04$. Bottom right: $T=0.06$.

%
%
{\bf FIGURE 9}. (Color online). 2D GEM4 phase diagram for low temperatures and pressures. The blue dots delimit the low-pressure boundary of the 2-cluster crystal phase (they were obtained by numerical free-energy calculations of the type described in the text). The blue dot at $T=0.04$ actually refers to the metastable equilibrium between the 2-phase and the ordinary-crystal phase, since the thermodynamically-stable phase is apparently the isotropic fluid. Note that the width of the hexatic region is actually smaller than indicated by the thickness of the melting (red) line.

%
%
{\bf FIGURE 10}. (Color online). $\rho$-$T$ phase diagram of the 2D GEM4 system. The hexatic-phase region is the narrow stripe delimited by the continuous freezing and triangular-crystal (TC) melting lines. The topology of the multiple-coexistence region is puzzling to us (hence we put a question mark): much more computational effort would be needed to resolve this region in full detail. The dotted lines are only schematic. The two isolated black crosses mark the $T=0.04$ coexistence densities of the TC and the 2CC with $n_c=1.8$. The 2CC coexisting with the TC at low temperature has $n_c=2$ (blue dots), whereas the 2CC coexisting with the fluid (f) has $n_c=1.9$ (red dots).

%
%
{\bf FIGURE 11}. (Color online). Specific energy $e$ (top) and number density $\rho$ (bottom) of the 2D GEM4 system along the $P=0.7$ isobar. Dots and triangles refer to data collected along the cooling and the heating trajectory, respectively. The data points are also made distinct for the number $M$ of MC sweeps carried out in each production run. Open symbols: $N=1152$ and $M=5\times 10^5$; full black symbols: $N=1152$ and $M=2\times 10^6$; full red symbols: $N=2688$ and $M=2\times 10^6$.

%
%
{\bf FIGURE 12}. (Color online). Thermal evolution of the RDF for small $r$, along the crystal-like $P=0.7$ path. The RDF profiles for several $T$ values are shown: 0.03-0.032 (black), 0.033-0.036 (blue), 0.037-0.040 (cyan), 0.041 (green), 0.042 (magenta), 0.043-0.045 (red). In the inset, the value of $n_c$ (as computed through Eq.\,(\ref{4-1})) is reported along the same path.

%
%
{\bf FIGURE 13}. (Color online). Isobaric melting of the cluster crystal with an average $n_c=1.9$ particles per site ($P=0.9$). We show energy and density data for two distinct system sizes ($N=1276$, open symbols; $N=5107$, full symbols). The straight-line segments across the data points merely serve as a guide to the eye. The crystal-like (red triangles) and fluid-like trajectories (blue dots) undergo a rather abrupt jump at different temperatures (hysteretic behavior), thus suggesting a first-order melting transition. The range of superheating is smaller for the larger sample, due to a longer length of the runs (see text).

%
%
{\bf FIGURE 14}. (Color online). Orientational correlation function $h_6(r)$ of the 2D GEM4 system in the solid-liquid transition region for $P=0.9$ along the crystal-like trajectory (left: $N=1276$; right: $N=5107$). In order to compute the OCF for a cluster crystal, a fake system of $N'$ particles is considered (with $N'<N$, depending on the system configuration) where each group of clustered particles is replaced with its center of mass while isolated particles are left untouched. The reported OCF profiles refer to the same temperature values for which Fig.\,13 shows data for the specific energy and the density (in all panels $T^*$ denotes the reduced temperature). Top panels: log-log scale; bottom panels: log-linear scale. The slope of the black dashed curve corresponds to a hypothetical $r^{-1/4}$ decay of $h_6(r)$. As $T$ increases, the large-distance behavior of $h_6(r)$ changes abruptly from constant (triangular solid) to exponential decay (isotropic fluid). The recovery of orientational correlations near half of the simulation-box length is a spurious effect due to the use of periodic boundary conditions.

\newpage
%
%
\begin{figure}
\centering
\includegraphics[width=16cm]{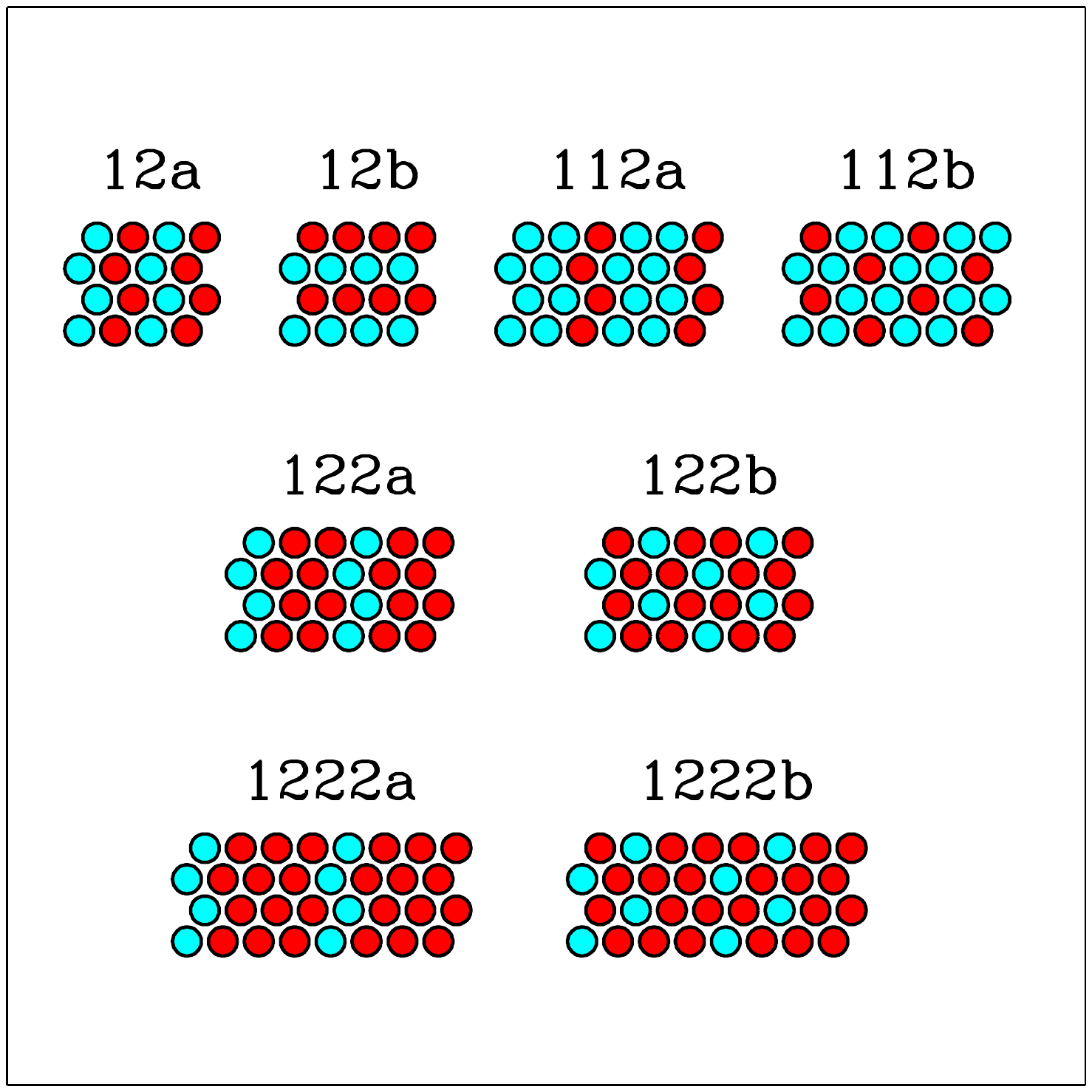}
\caption{
}
\label{fig1}
\end{figure}

%
%
\begin{figure}
\centering
\includegraphics[width=16cm]{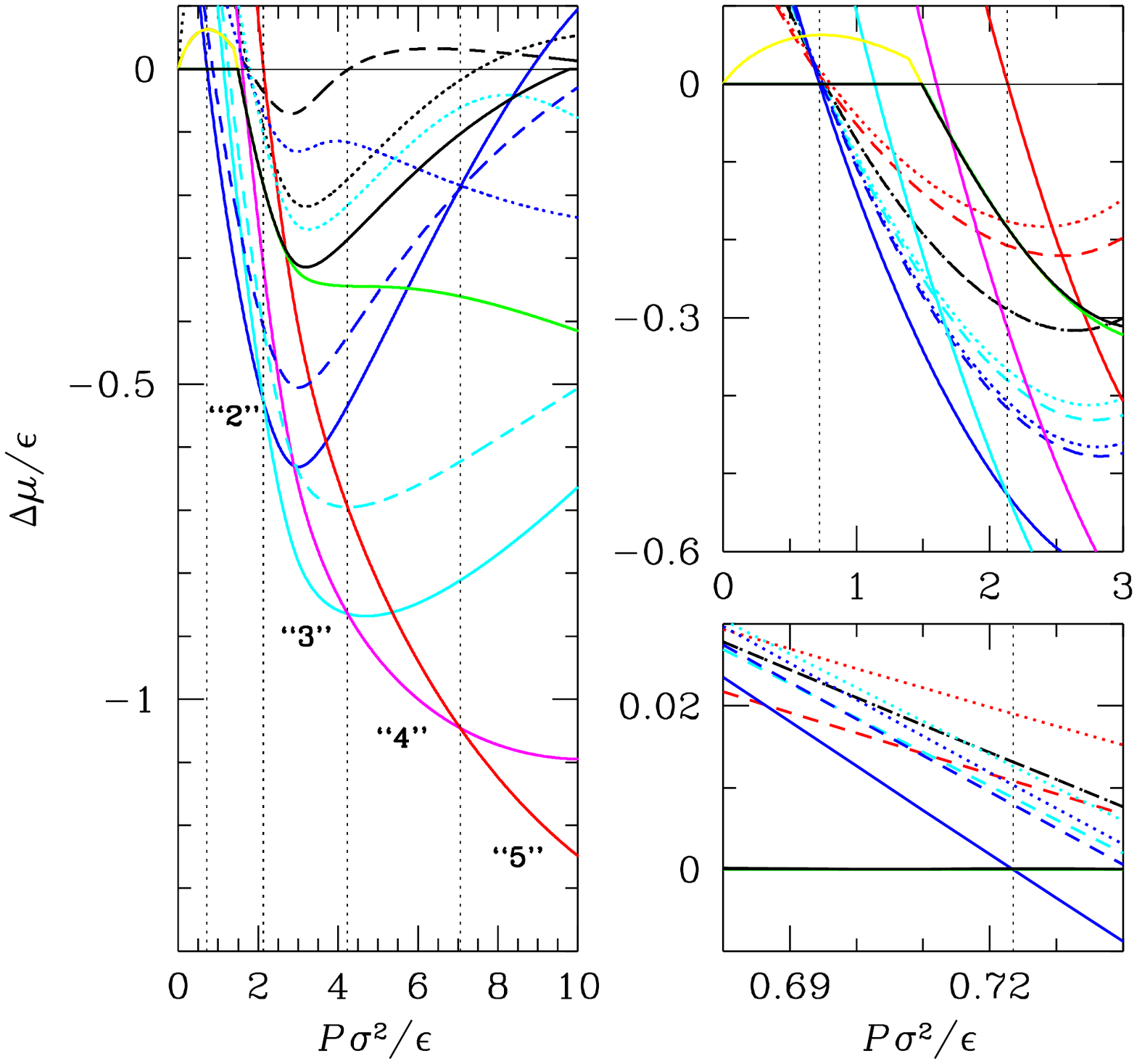}
\caption{
}
\label{fig2}
\end{figure}

%
%
\begin{figure}
\centering
\includegraphics[width=16cm]{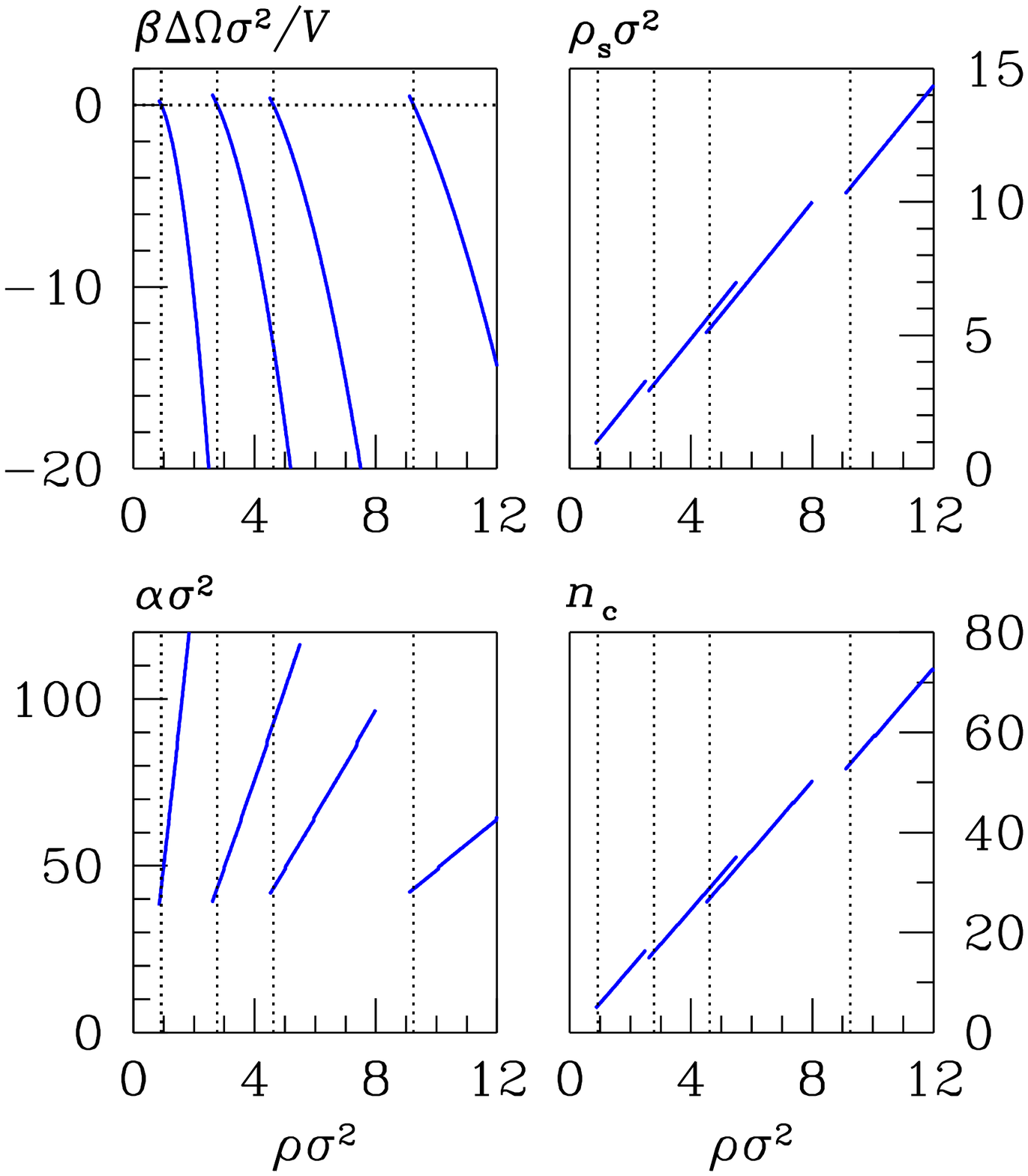}
\caption{
}
\label{fig3}
\end{figure}

%
%
\begin{figure}
\centering
\includegraphics[width=16cm]{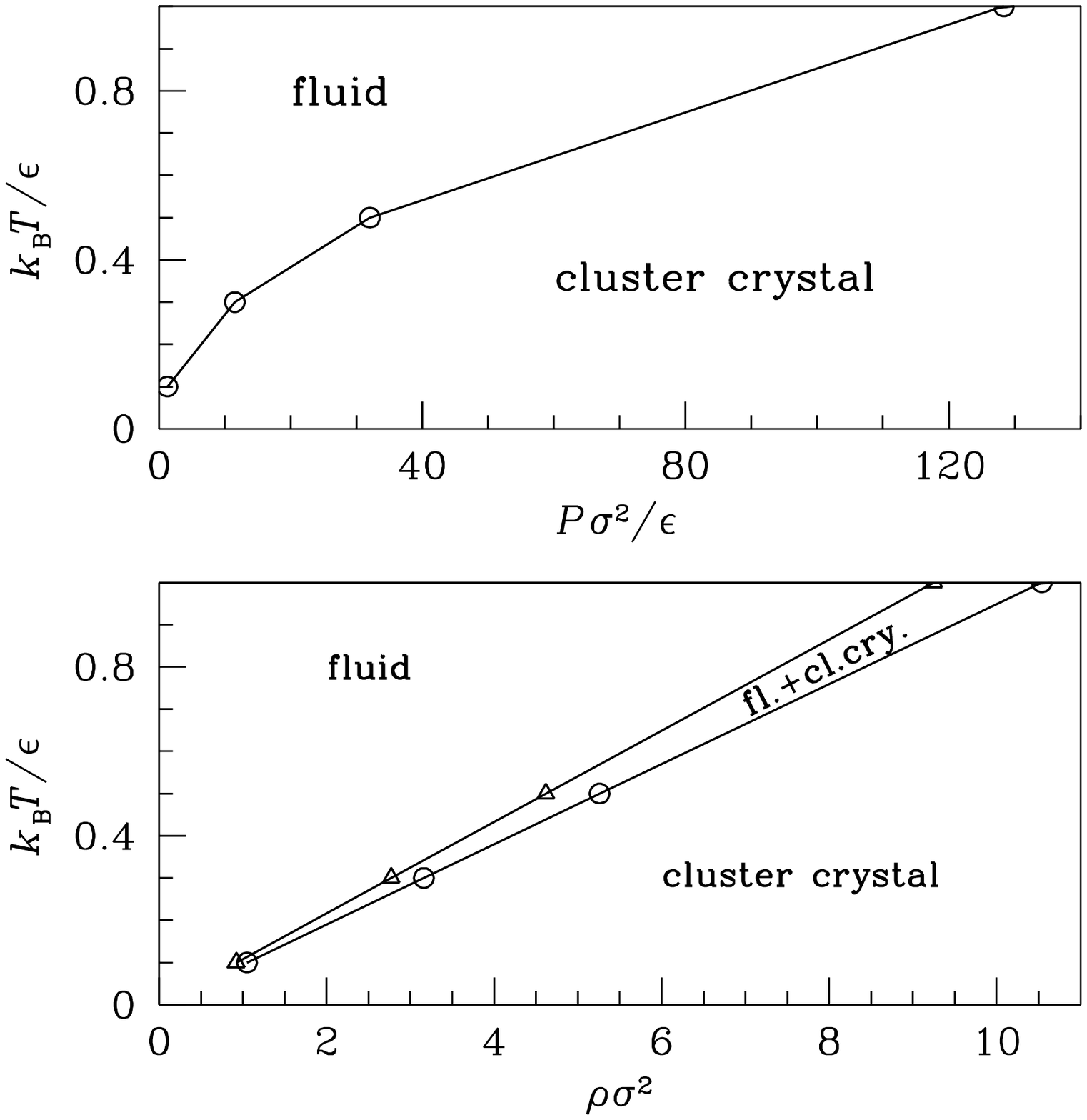}
\caption{
}
\label{fig4}
\end{figure}

%
%
\begin{figure}
\centering
\includegraphics[width=16cm]{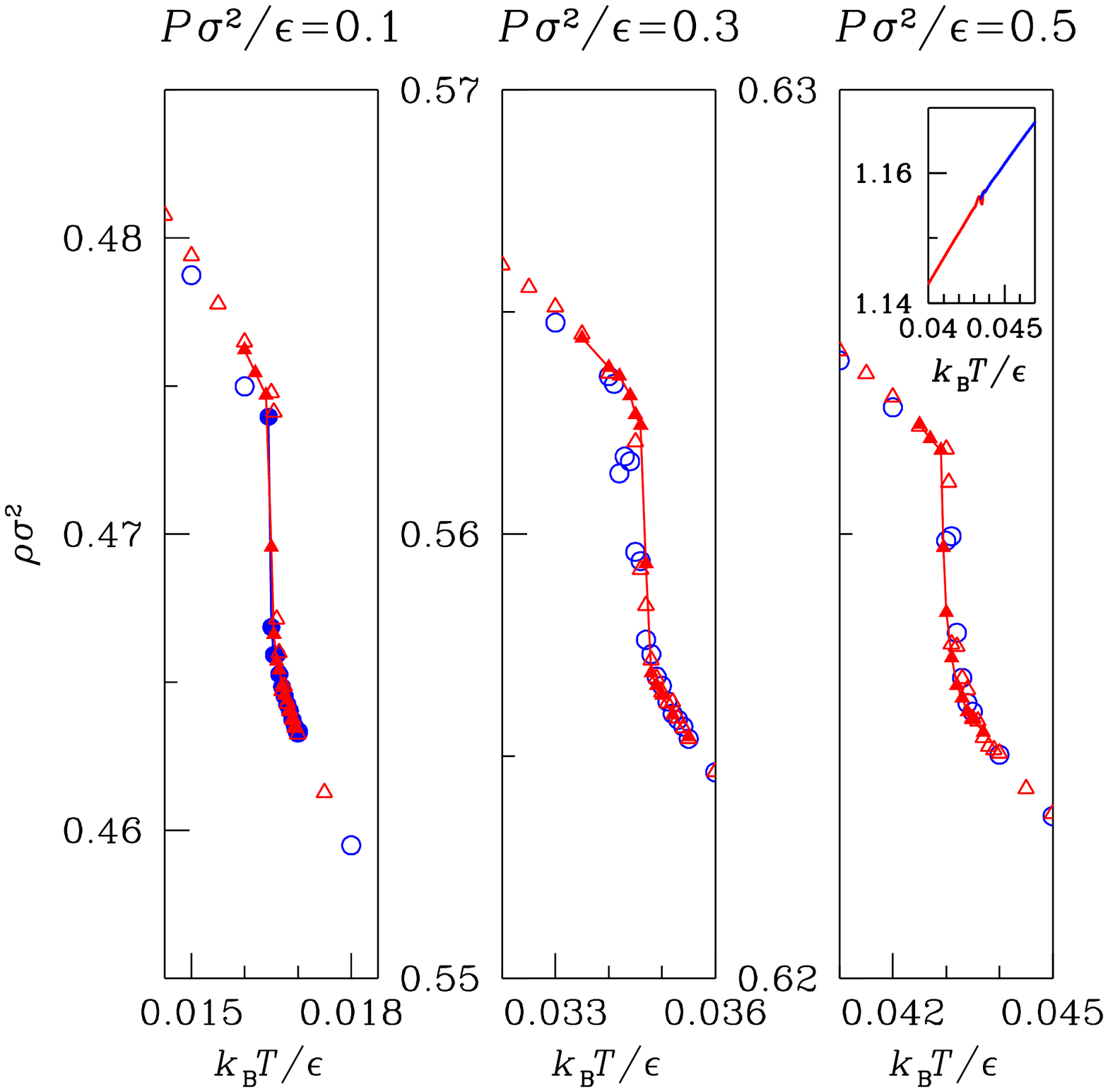}
\caption{
}
\label{fig5}
\end{figure}

%
%
\begin{figure}
\centering
\includegraphics[width=16cm]{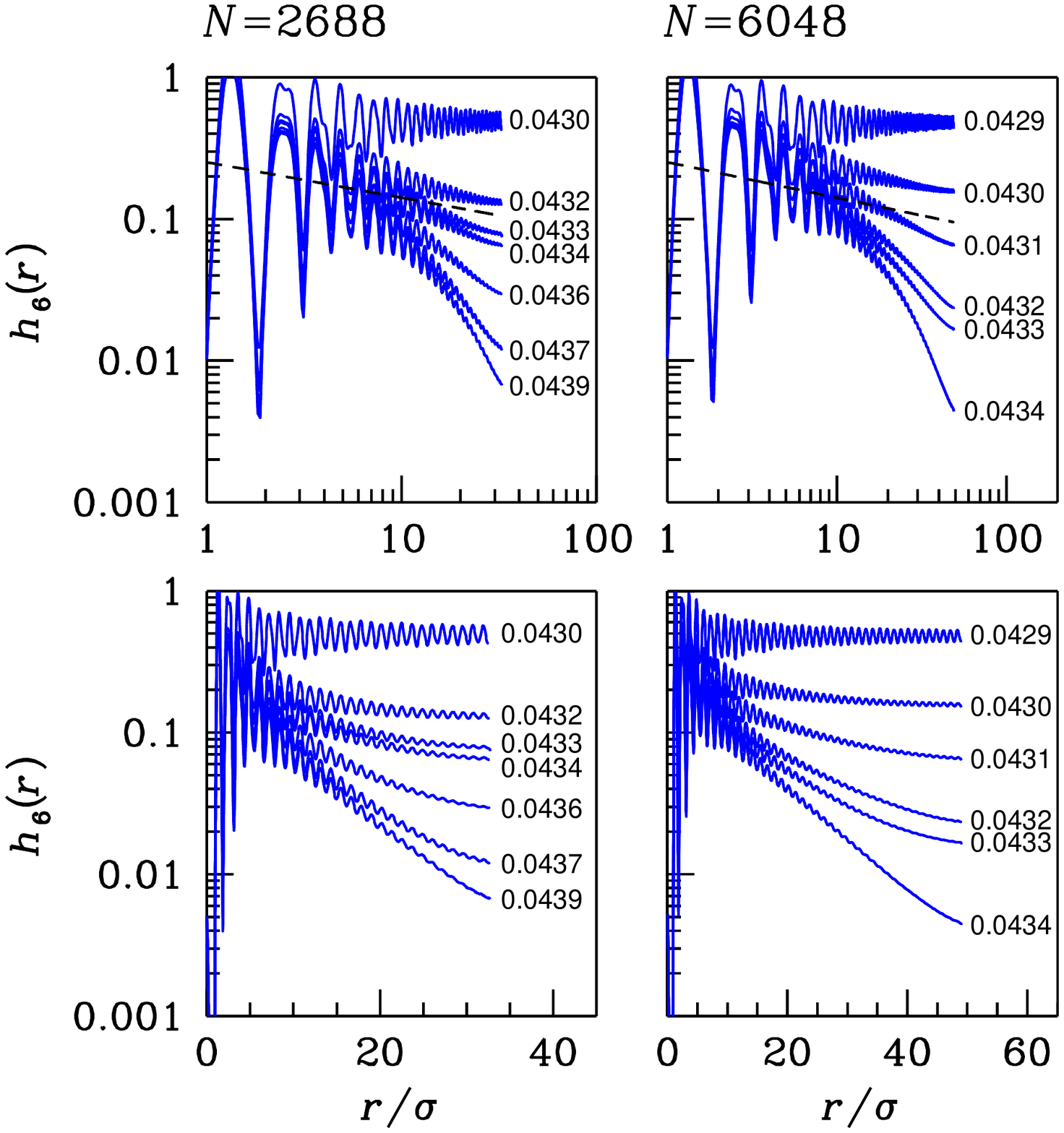}
\caption{
}
\label{fig6}
\end{figure}

%
%
\begin{figure}
\centering
\includegraphics[width=16cm]{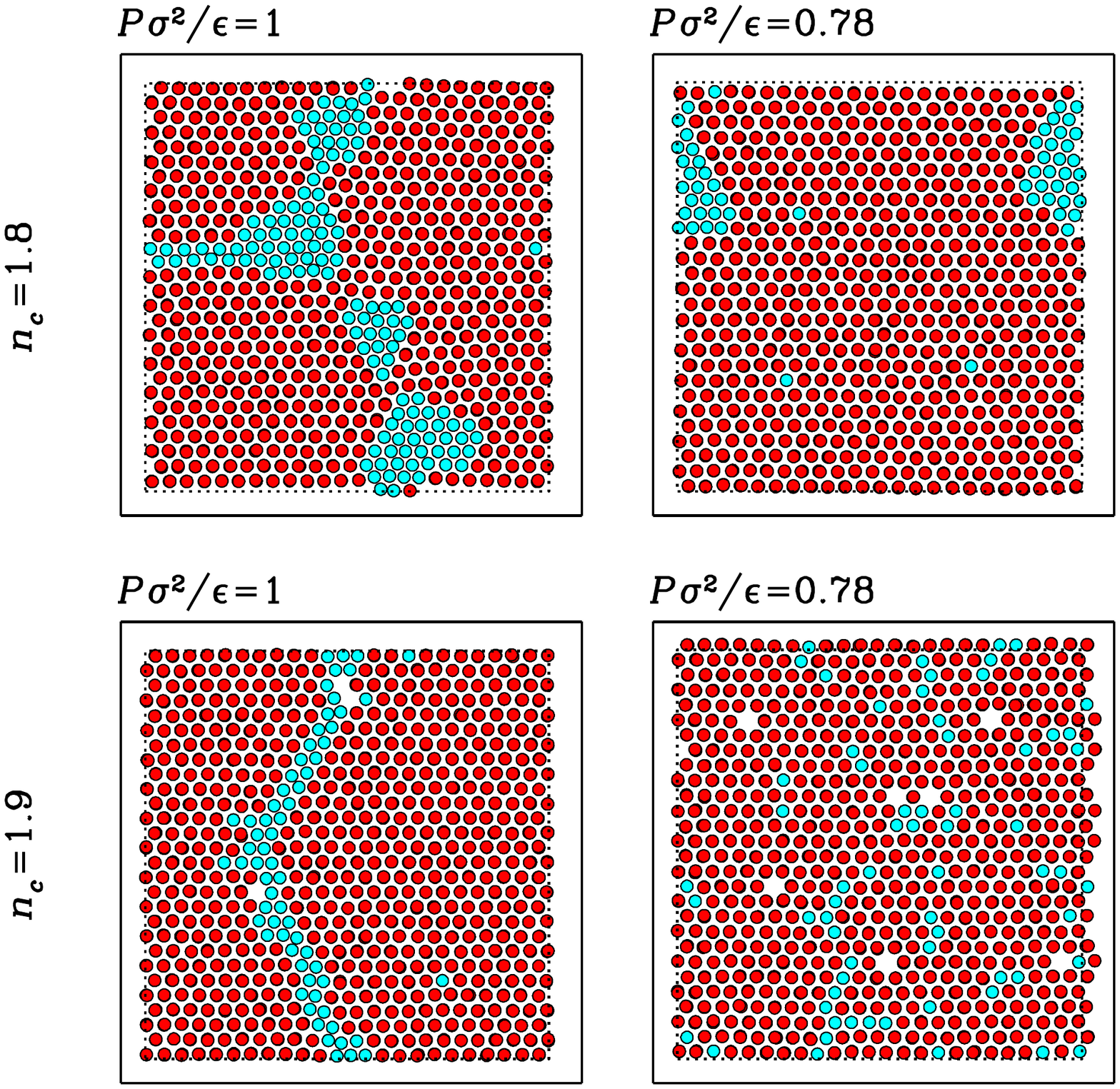}
\caption{
}
\label{fig7}
\end{figure}

%
%
\begin{figure}
\centering
\includegraphics[width=16cm]{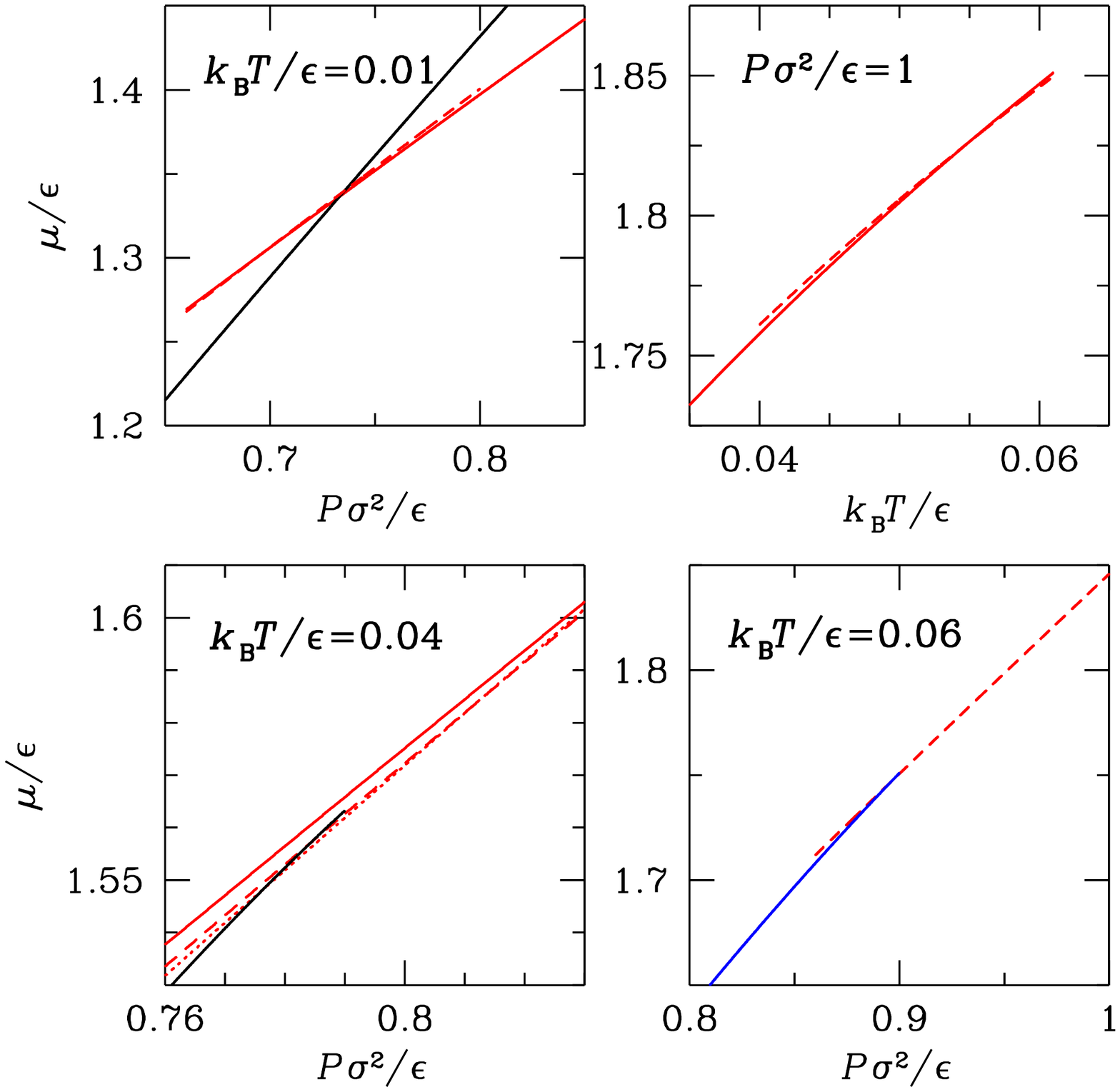}
\caption{
}
\label{fig8}
\end{figure}

%
%
\begin{figure}
\centering
\includegraphics[width=16cm]{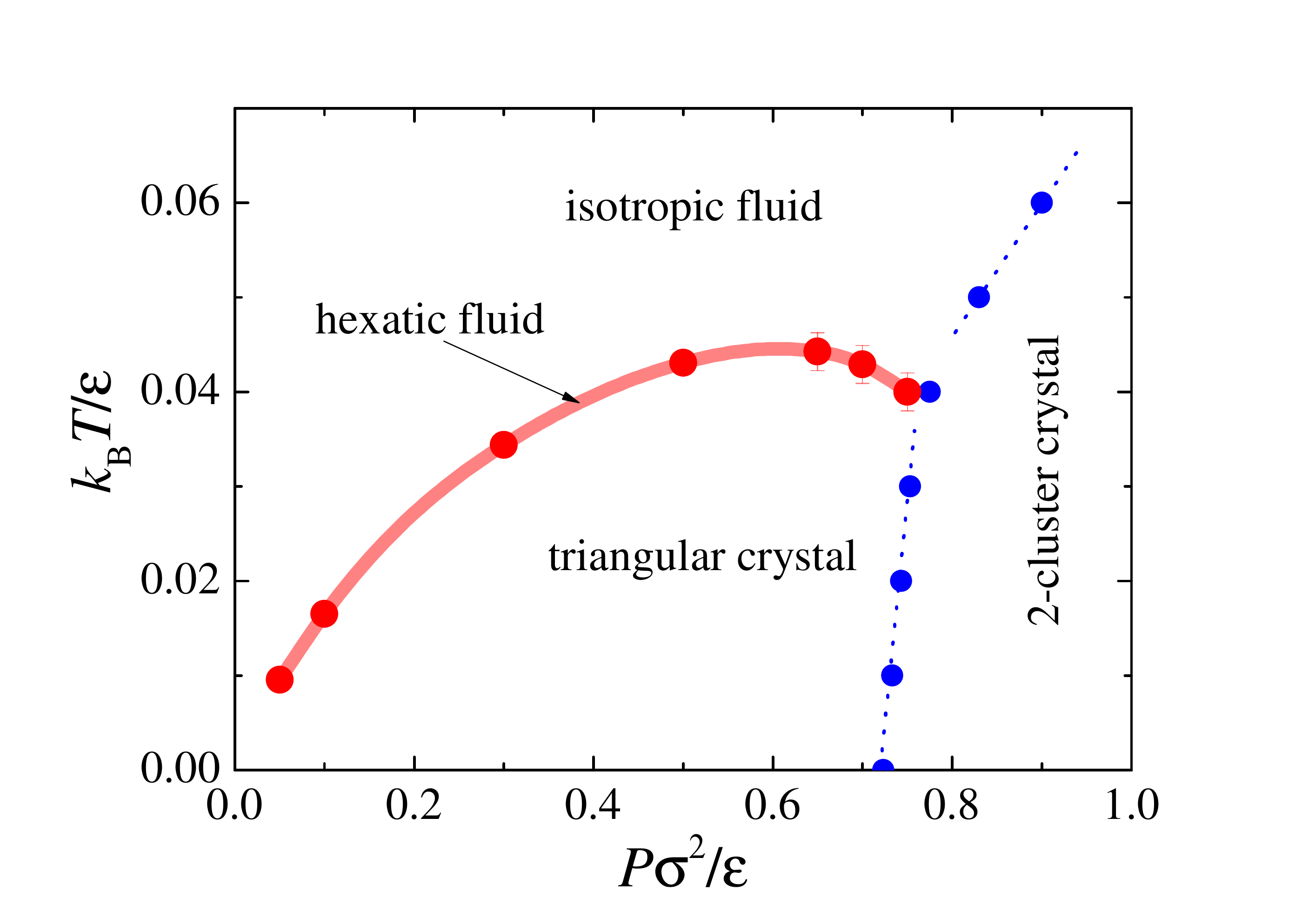}
\caption{
}
\label{fig9}
\end{figure}

%
%
\begin{figure}
\centering
\includegraphics[width=16cm]{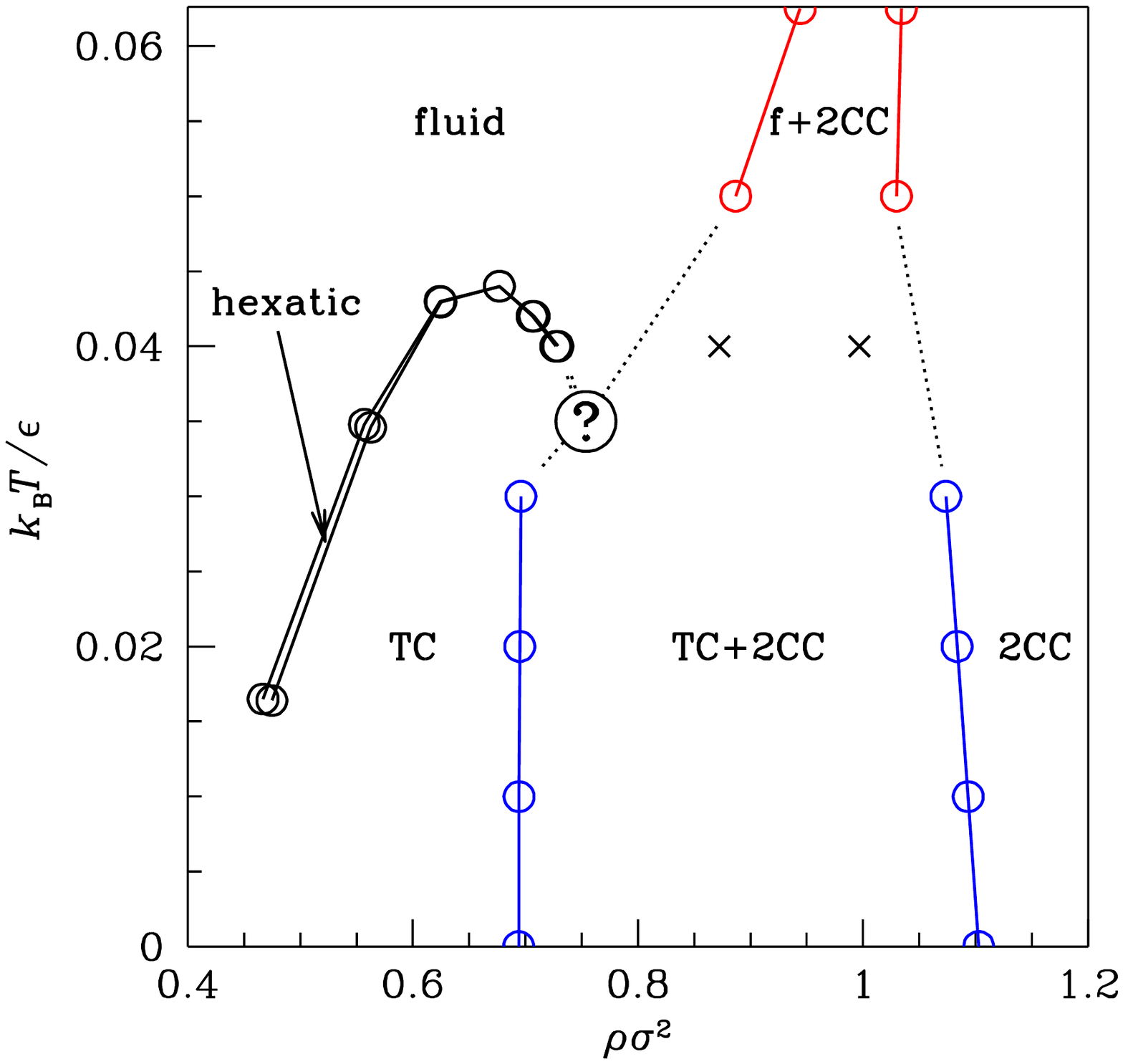}
\caption{
}
\label{fig10}
\end{figure}

%
%
\begin{figure}
\centering
\includegraphics[width=16cm]{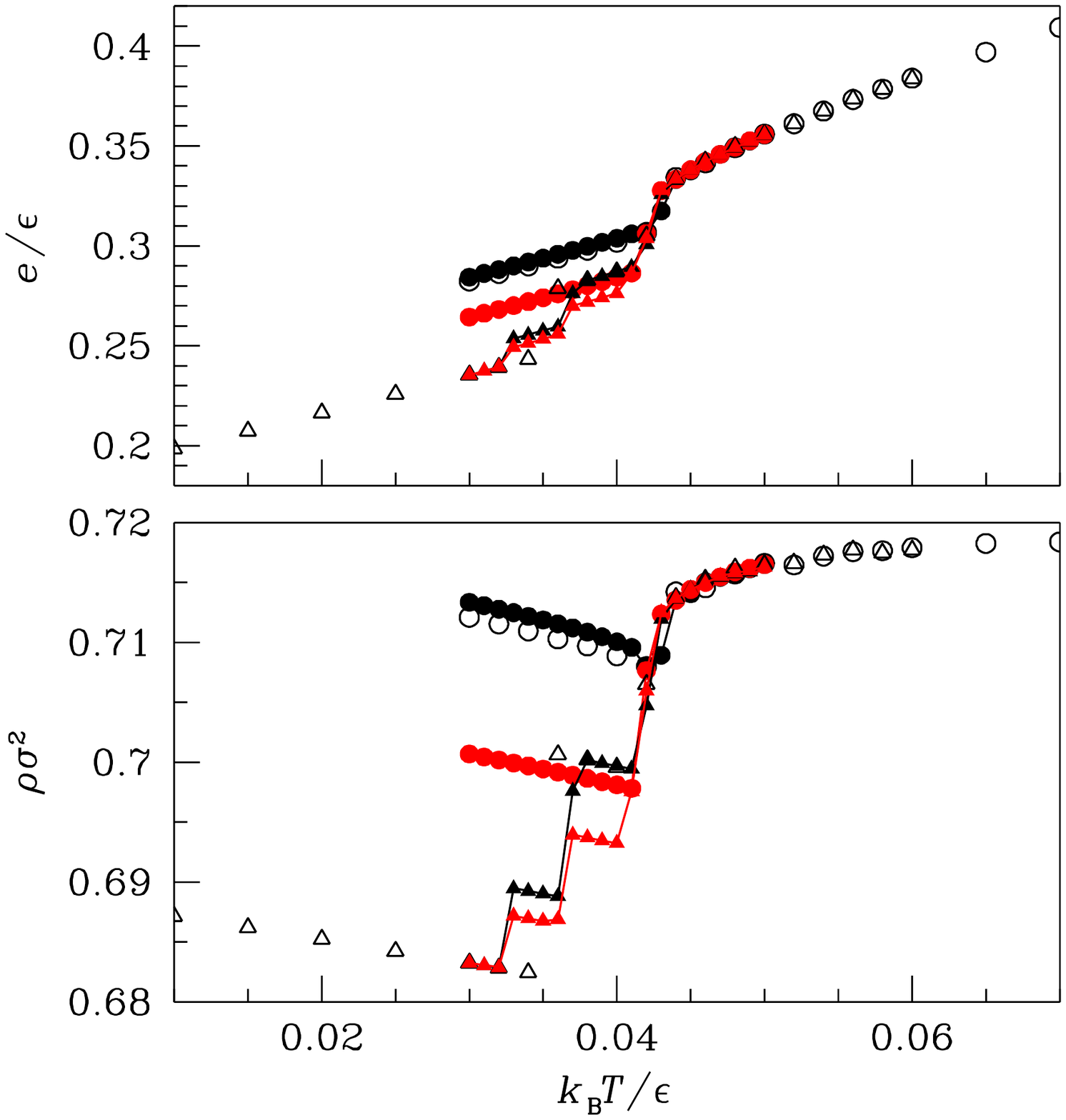}
\caption{
}
\label{fig11}
\end{figure}

%
%
\begin{figure}
\centering
\includegraphics[width=16cm]{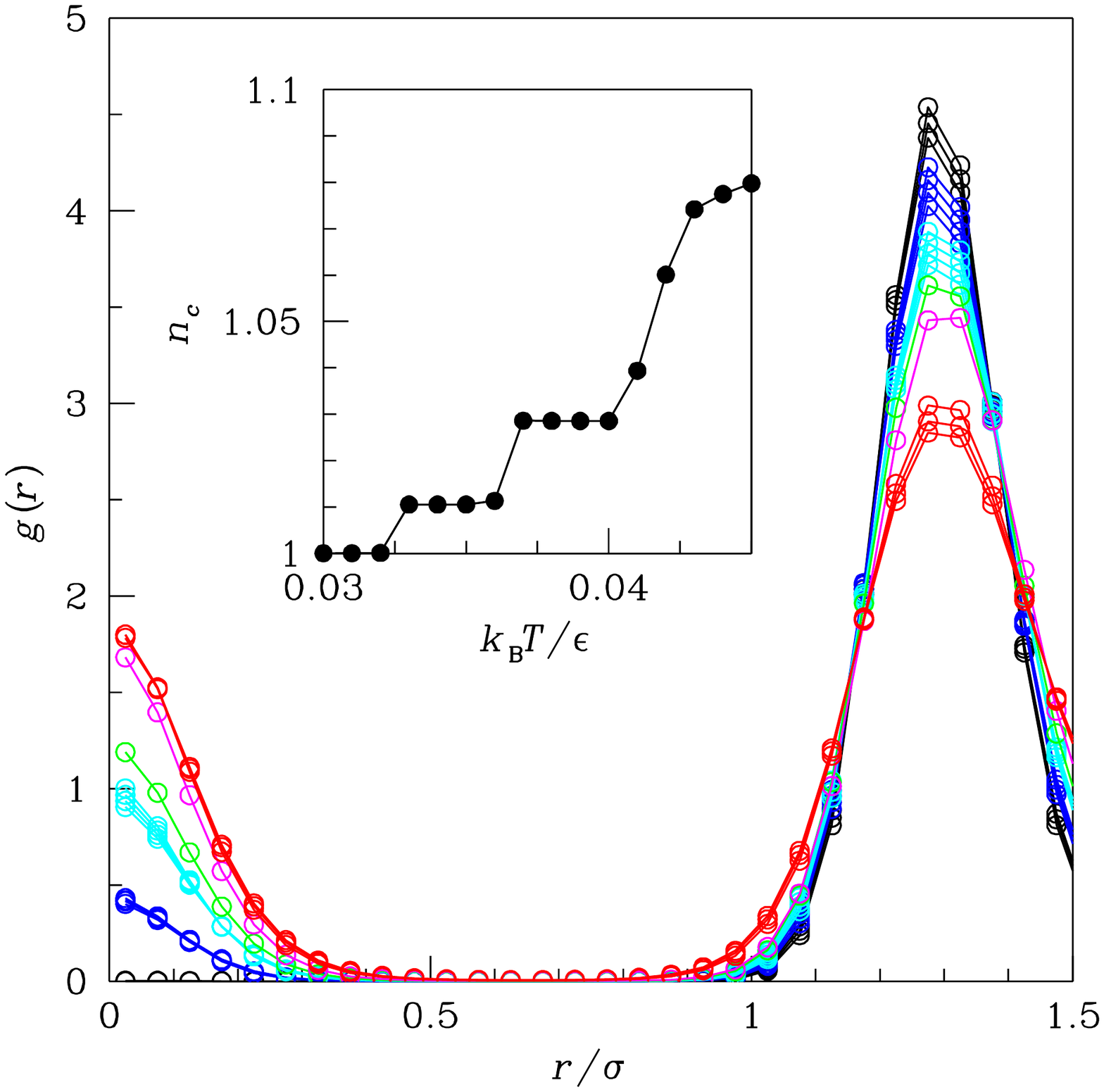}
\caption{
}
\label{fig12}
\end{figure}

%
%
\begin{figure}
\centering
\includegraphics[width=16cm]{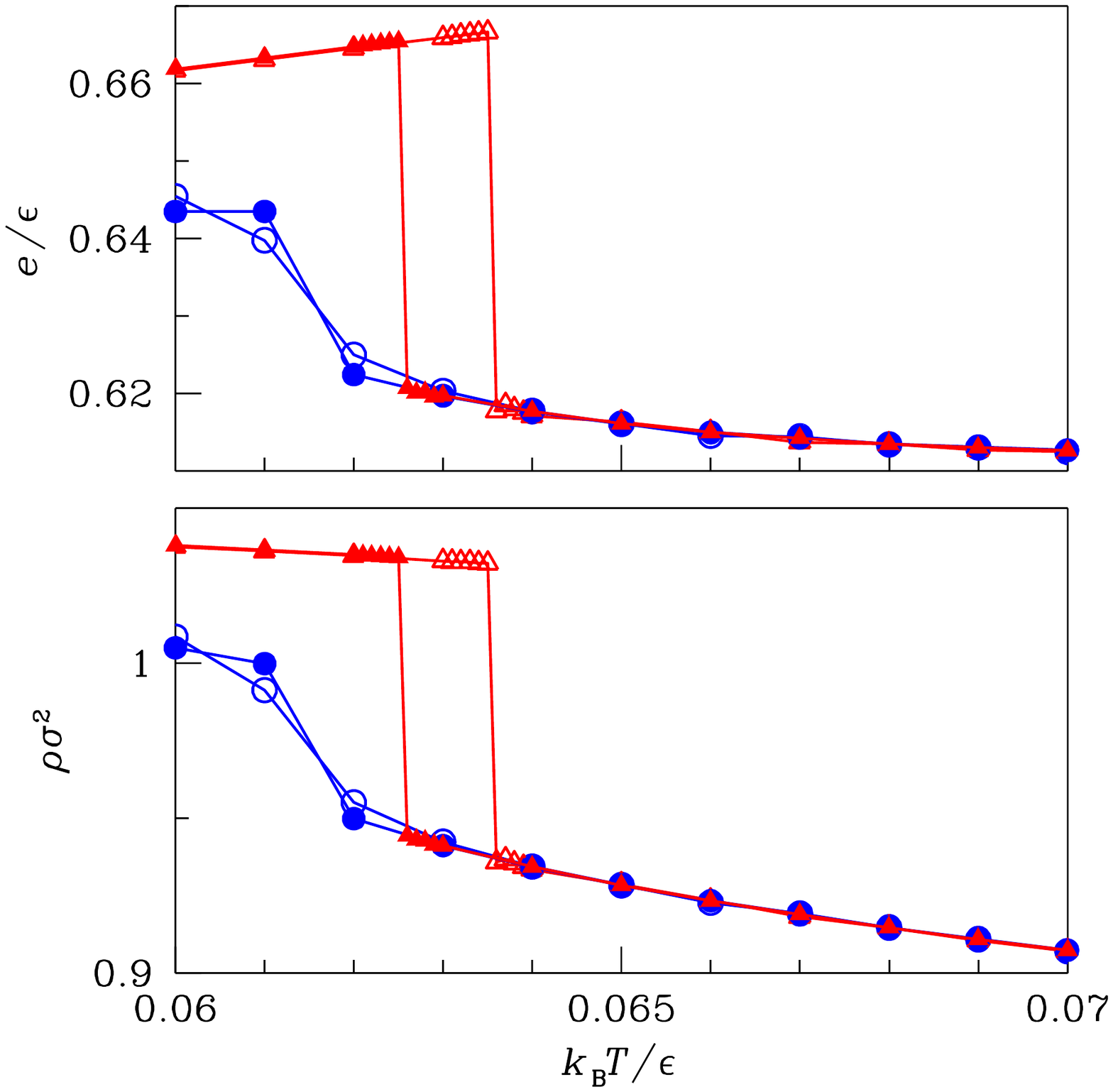}
\caption{
}
\label{fig13}
\end{figure}

%
%
\begin{figure}
\centering
\includegraphics[width=16cm]{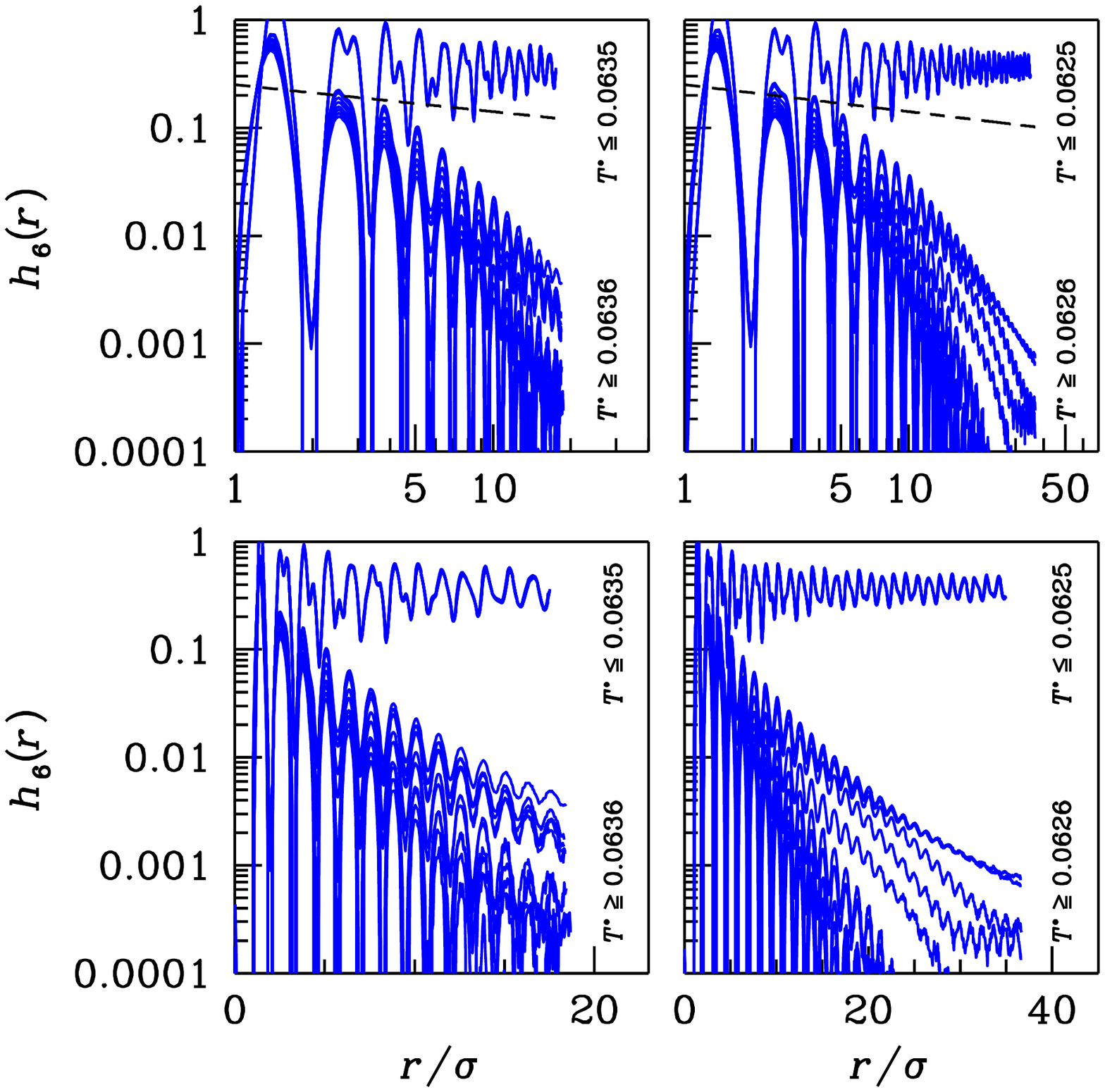}
\caption{
}
\label{fig14}
\end{figure}
\end{document}